\documentclass[letter,11pt,DIV=12,abstract=true,numbers=noenddot,titlepage=false,twocolumn=false,draft=false]{scrartcl}
\pdfoutput=1 
\usepackage{geometry}
\makeatletter
\DeclareOldFontCommand{\rm}{\normalfont\rmfamily}{\mathrm}
\DeclareOldFontCommand{\sf}{\normalfont\sffamily}{\mathsf}
\DeclareOldFontCommand{\tt}{\normalfont\ttfamily}{\mathtt}
\DeclareOldFontCommand{\bf}{\normalfont\bfseries}{\mathbf}
\DeclareOldFontCommand{\it}{\normalfont\itshape}{\mathit}
\DeclareOldFontCommand{\sl}{\normalfont\slshape}{\@nomath\sl}

\usepackage[usenames,dvipsnames]{color}
  \definecolor{hgreen}{rgb}{0,.3,0}
  \definecolor{hred}{rgb}{.3,0,0}
  \definecolor{hblue}{rgb}{0,0,.3}
  \definecolor{LightGray}{gray}{0.95}
  \definecolor{gray}{gray}{0.6}
\usepackage[
	    colorlinks=true,
	    linkcolor=hblue,
	    citecolor=hgreen,
	    filecolor=hblue,
	    urlcolor=hred
	    ]{hyperref}
\usepackage{mathrsfs}
\usepackage{amsmath}
\usepackage{amssymb}
\usepackage{slashed}
\usepackage[titletoc,title]{appendix}
\usepackage[affil-it]{authblk}
\usepackage{libertine}
\usepackage[numbers,sort&compress]{natbib}
\usepackage{graphicx}

\numberwithin{equation}{section}

\newcommand{\Lag}{\mathscr{L}}

\newcommand{\xhw}{x_{hW}}
\newcommand{\xth}{x_{th}}
\newcommand{\xtz}{x_{tZ}}
\newcommand{\xwh}{x_{Wh}}
\newcommand{\xhz}{x_{hZ}}
\newcommand{\xzh}{x_{Zh}}

\definecolor{Blu}{rgb}{0.,0.,1.}
\definecolor{Red}{rgb}{1.,0.,0.}

\begin{document}
\renewcommand\Authands{, }
\unitlength = 1mm

\title{\boldmath Electric dipole moment constraints on CP-violating light-quark Yukawas}
\subtitle{\flushright \vspace{-25ex} \rm DO-TH 18/26 \vspace{20ex}}
\date{\today}
\author[a]{Joachim Brod%
        \thanks{\texttt{joachim.brod@uc.edu}}}
\author[b]{Dimitrios Skodras%
        \thanks{\texttt{dimitrios.skodras@tu-dortmund.de}}}
\affil[a]{{\large Department of Physics, University of Cincinnati, Cincinnati, OH 45221, USA}}
\affil[b]{{\large Fakult\"at f\"ur Physik, TU Dortmund, D-44221 Dortmund, Germany}}

\maketitle

\begin{abstract}
Nonstandard CP violation in the Higgs sector can play an essential
role in electroweak baryogenesis. We calculate the full two-loop
matching conditions of the standard model, with Higgs Yukawa couplings
to light quarks modified to include arbitrary CP-violating phases,
onto an effective Lagrangian comprising CP-odd electric and
chromoelectric light-quark (up, down, and strange) dipole operators.
We find large isospin-breaking contributions of the electroweak
diagrams. Using these results, we obtain significant constraints on
the phases of the light-quark Yukawas from experimental bounds on the
neutron and mercury electric dipole moments.
\end{abstract}
\setcounter{page}{1}

\section{Introduction}
\label{sec:introduction}

Electric dipole moments (EDMs) of atomic systems and elementary
particles are sensitive probes of CP violation~\cite{Pospelov:2005pr,
  Chupp:2017rkp, Yamanaka:2017mef}.  New sources of CP violation
beyond the Standard Model (SM) are a necessary ingredient of models
explaining the observed baryon asymmetry of the universe
(``baryogenesis''). A particularly appealing scenario is electroweak
baryogenesis (see Ref.~\cite{Morrissey:2012db} for a review), as it
can potentially be probed at the LHC.  In many models, electroweak
baryogenesis is driven by a CP-violating phase in the Higgs-top
coupling. This phase also induces an electric dipole moment (EDM) in
various elementary particles and hadronic systems.  Hence,
experimental bounds on EDMs, for instance of the neutron, give strong
constraints on new phases in the top Yukawa and naively exclude many
models of baryogenesis.

On the other hand, as pointed out in Ref.~\cite{Huber:2006ri}, phases
in Yukawas other than the top Yukawa are barely relevant for
electroweak baryogenesis, whereas they can lead to a substantial
modification of the EDM bounds. This motivates a detailed study of
CP-violating contributions to {\em all} Yukawa couplings.

In Ref.~\cite{Brod:2013cka}, EDM constraints on the Yukawa couplings
of the third fermion generation (top, bottom, tau) were studied. The
remaining large theory uncertainty in the constraint for CP violation
in the bottom Yukawa was addressed in Ref.~\cite{Brod:2018pli}, and the
analysis extended to include also the charm quark. EDM constraints on
the electron Yukawa were obtained in
Ref.~\cite{Altmannshofer:2015qra}.

The constraints on light-quark Yukawas (up, down, and strange) were
studied in Ref.~\cite{Chien:2015xha} in the context of effective
dimension-six Higgs--quark interactions. In that work, a fit was
performed to one or two Yukawa couplings at a time, using hadronic
EDMs induced by Barr-Zee diagrams with virtual top quarks in the
loop. In the present publication, we calculate the full set of
contributing two-loop diagrams induced by CP-violating phases in the
light-quark Yukawas. In particular, we show that the contribution of
bosonic diagrams dominates over the top-loop diagrams. Further, we
show that the particular pattern of relative contributions to the
electric and chromoelectric dipole operators at low energies retains
the large complementarity of constraints from the neutron and mercury
EDMs.

This article is organized as follows. In Sec.~\ref{sec:calc} we
present the framework of our calculation, as well as the full analytic
results, and the renormalization-group (RG) evolution from the
electroweak to the hadronic scale. In Sec.~\ref{sec:num} we study the
numerical implications of our results for CP violation in the
light-quark Yukawas, and we conclude in Sec.~\ref{sec:conclusions}.

\section{Setup and calculation}
\label{sec:calc}

Our starting point is the SM Lagrangian with modified Yukawa couplings
of the form
\begin{align}
\label{eq:LagHqq}
\Lag_{h-q-q}  &=  - \frac{y^{\text{SM}}_q}{\sqrt{2}} \kappa_q \bar
q\left( \cos\phi_q + i\gamma_5\sin\phi_q \right)q\,h\,, 
\end{align}
where $h$ denotes the SM Higgs field in the broken phase, and $q$ the
SM quark fields. Moreover, $y^{\text{SM}}_q \equiv
e\,m_q/(\sqrt{2}s_wM_W)$ is the SM Yukawa, with $e$ the positron
charge, $s_w$ the sine of the weak mixing angle, and $M_W$ the
$W$-boson mass, respectively. The mass of the light quark $q$ is
denoted by $m_q$. The real parameters $\kappa_q\geq0$ parameterize
modifications to the absolute value of the SM Yukawa couplings, while
the phases $\phi_q \in [0,2\pi)$ parameterize CP violation and the sign
  of the Yukawa. The SM value is obtained in the limit $\kappa_q=1$
  and $\phi_q=0$.

In this work we are interested in constraining the phases of the
light-quark Yukawas, $q=u,d,s$, via their contributions to hadronic
EDMs. They are induced by the partonic low-energy effective
Lagrangian~\cite{Engel:2013lsa} valid at energy scales of the order of
one GeV,
\begin{equation} \label{eq:LeffN}
{\cal L}_{\rm eff} = - d_q \, \frac{i}{2} \, \bar q \sigma^{\mu\nu}
\gamma_5 q \, F_{\mu\nu} - \tilde d_q \, \frac{ig_s}{2} \, \bar q
\sigma^{\mu\nu} T^a \gamma_5 q \, G_{\mu\nu}^a \,,
\end{equation}
with $\sigma^{\mu\nu} = \tfrac{i}{2} [\gamma^\mu, \gamma^\nu]$.

We calculate the contributions to this Lagrangian of the modified
Yukawa couplings in Eq.~\eqref{eq:LagHqq} by first matching the
modified SM to the effective Lagrangian
\begin{equation} \label{eq:Leff}
\Lag_\text{eff} = - \sqrt{2} G_F \, \sum_q \sum_{i=1,2} C_i^q O_i^q +
\ldots\,
\end{equation}
obtained by integrating out the heavy gauge bosons, the top quark, and
the Higgs boson at the weak scale $\mu_\text{ew} \sim M_h =
125.18\,$GeV. Here, the sum runs over all active quark fields below
the weak scale ($q= u,d,s,c,b$), and the operators are defined as
\begin{align} \label{eq:dipole}
O_1^q & = \frac{ieQ_q}{2} \, m_q \, \bar q \sigma^{\mu \nu} \gamma_5 q \, F_{\mu \nu} \,,&
O_2^q & = -\frac{i}{2} \, g_s \, m_q \, \bar q \sigma^{\mu \nu} \gamma_5 T^a q \, G^a_{\mu \nu} \,. &
\end{align}
A plethora of other effective operators is generated by the matching
procedure, for instance, CP-odd four-fermion operators. These other
operators are either subleading or do not directly contribute to the
hadronic EMDs, and are denoted by the ellipsis in Eq.~\eqref{eq:Leff}.
%
\begin{figure}[t]
\begin{center}
\includegraphics[width=0.25\textwidth]{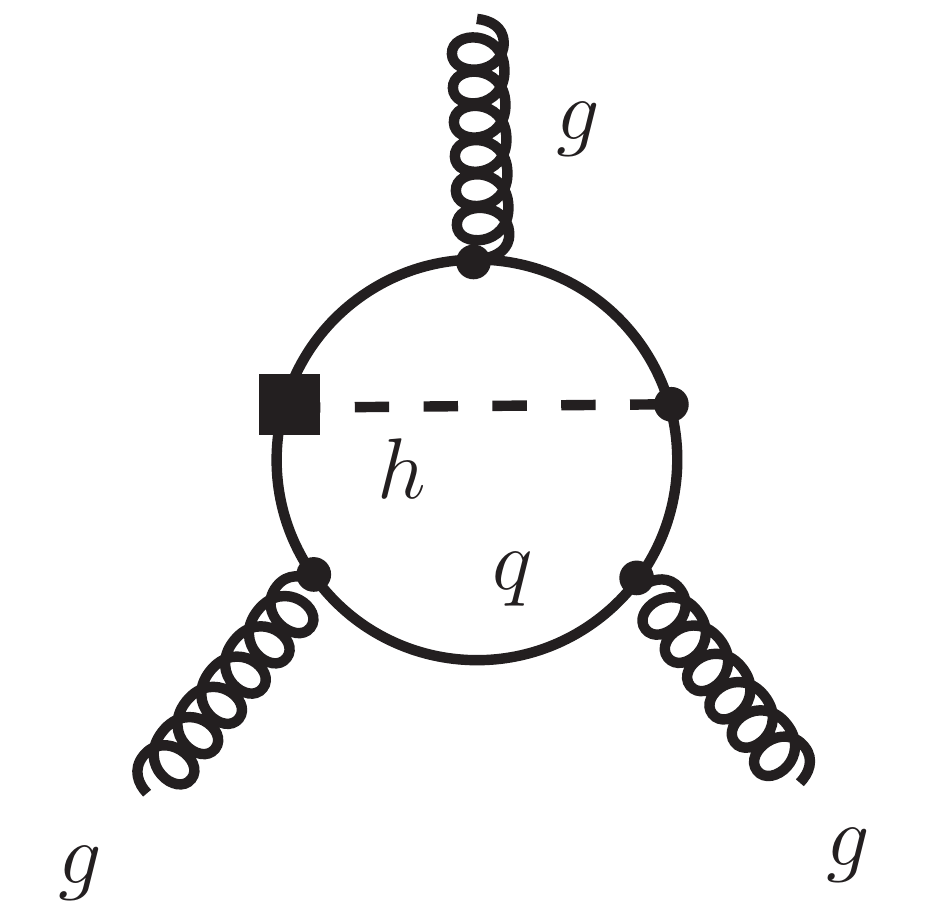}
\end{center}
\caption{The contributions to the Weinberg operator are suppressed
  with respect to the dipole contributions by an additional power of a
  light-quark Yukawa and are neglected. Here and in the following, the
  CP-violating Higgs coupling to the light quarks $q=u,d,s$ is denoted
  by a black square. \label{fig:2loopEDM:weinberg}}
\end{figure}
%
%
\begin{figure}[t]
\begin{center}
\includegraphics[width=0.3\textwidth]{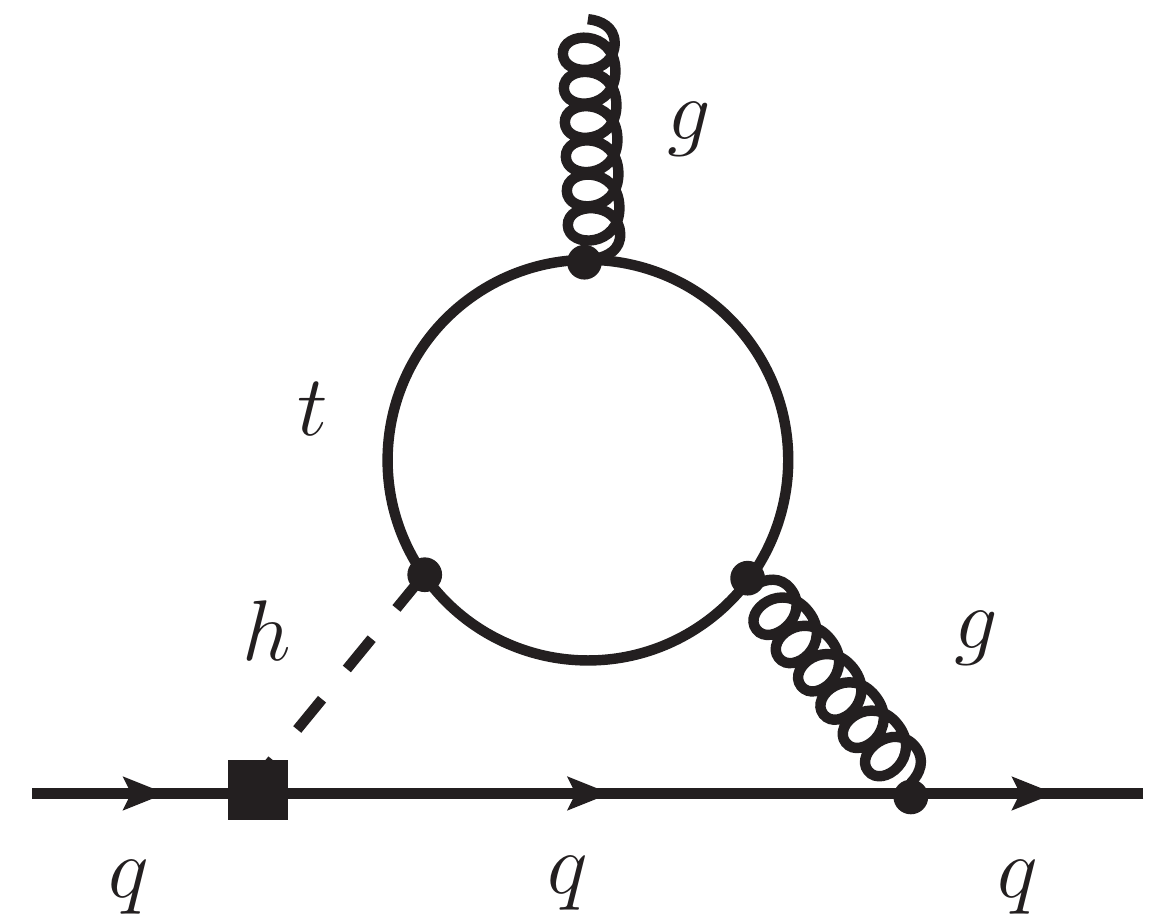}~~~~~~~~~
\includegraphics[width=0.3\textwidth]{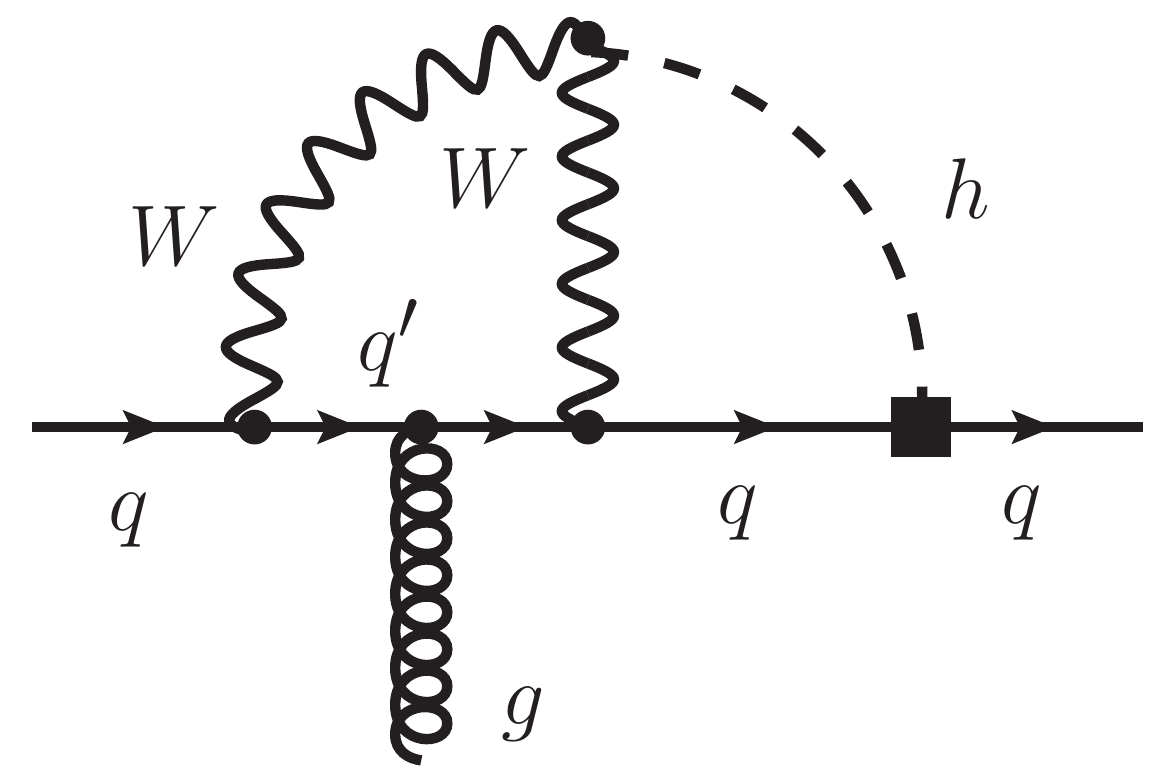}~~~~~~~~~
\includegraphics[width=0.3\textwidth]{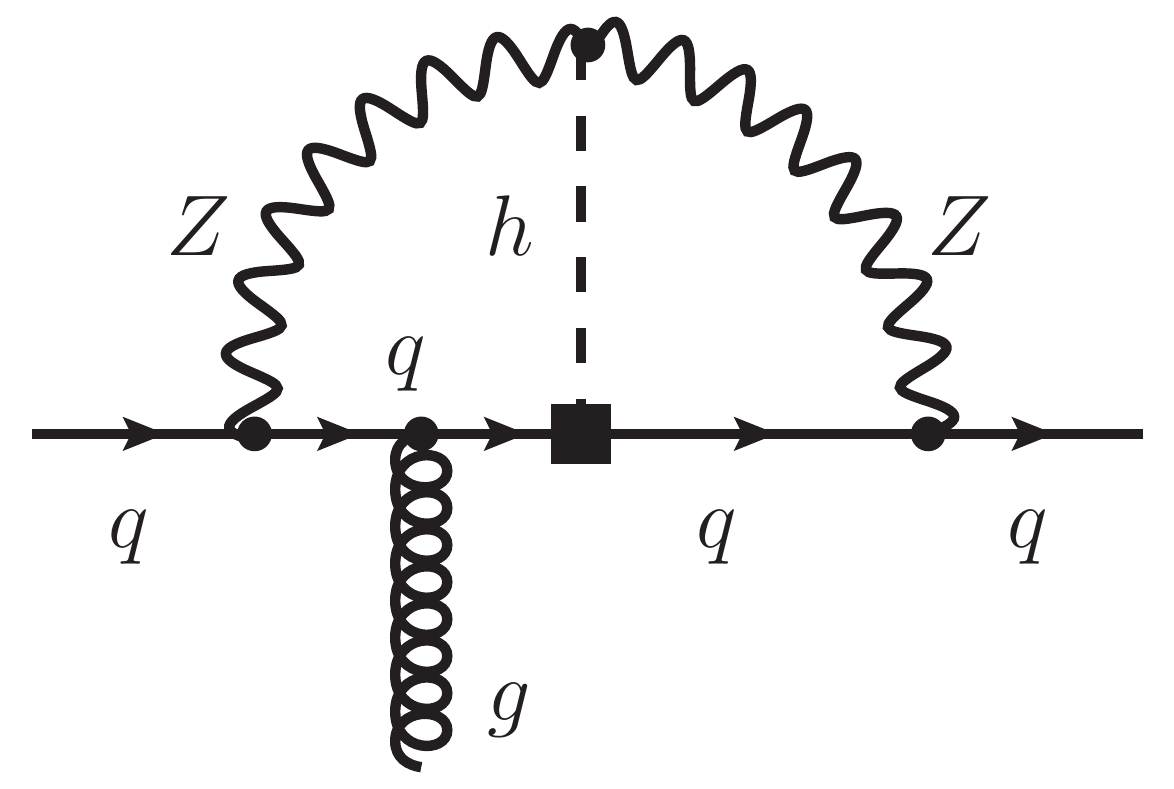}
\end{center}
\caption{Sample two-loop Feynman diagrams inducing an chromoelectric
  dipole moment for the light quark $q=u,d,s$ through a CP-violating
  Higgs coupling. The label $q'$ denotes a quark with opposite weak
  isospin with respect to~$q$. \label{fig:2loopEDM:gluonic}}
\end{figure}
%
As a notable example, we mention that the Weinberg
operator~\cite{Weinberg:1989dx}
\begin{equation} \label{eq:op:weinberg}
\begin{split}
O_3 & = -\frac{1}{3\,g_s} f^{abc} \, G_{\mu \sigma}^a G_{\nu}^{b, \sigma} \widetilde G^{c, \mu \nu} \,
\end{split}
\end{equation}
also receives a contribution from modified quark couplings (see
Fig.~\ref{fig:2loopEDM:weinberg}). However, since we are interested in
the light-quark Yukawa couplings, these contributions are suppressed
by an additional power of a small Yukawa coupling with respect to the
dipole operators~\eqref{eq:dipole}. Furthermore, the dipole operators
do not mix into the Weinberg operator, hence it plays no role in our
calculation.

\begin{figure}[t]
\begin{center}
\includegraphics[width=0.3\textwidth]{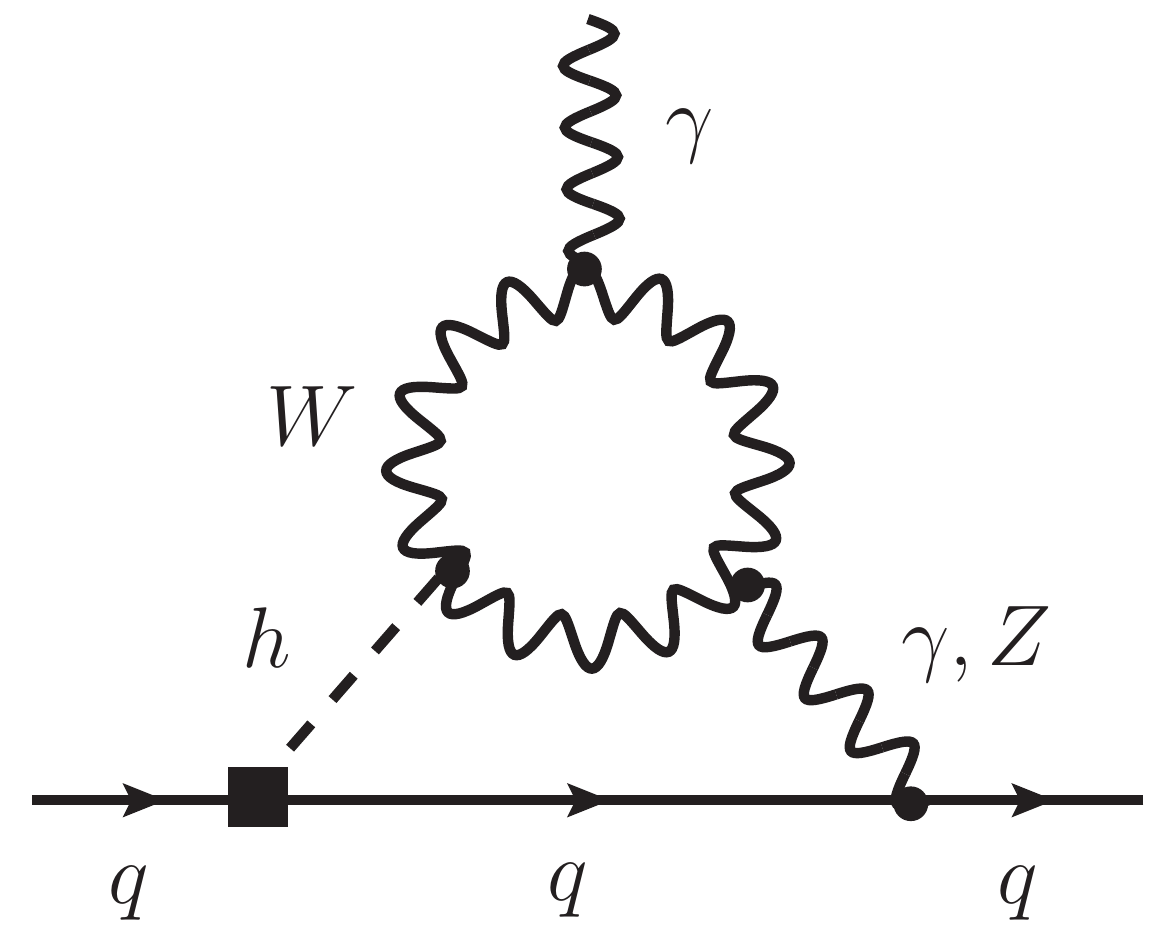}~~~~~
\includegraphics[width=0.3\textwidth]{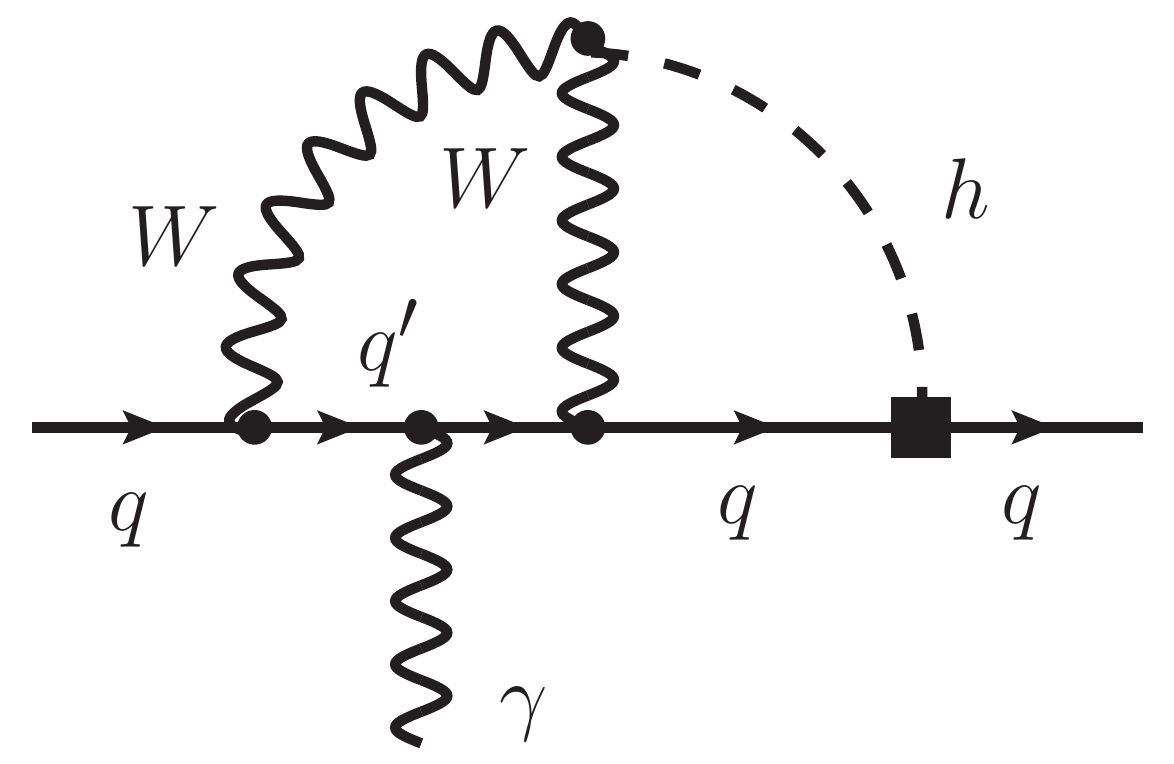}~~~~~
\includegraphics[width=0.3\textwidth]{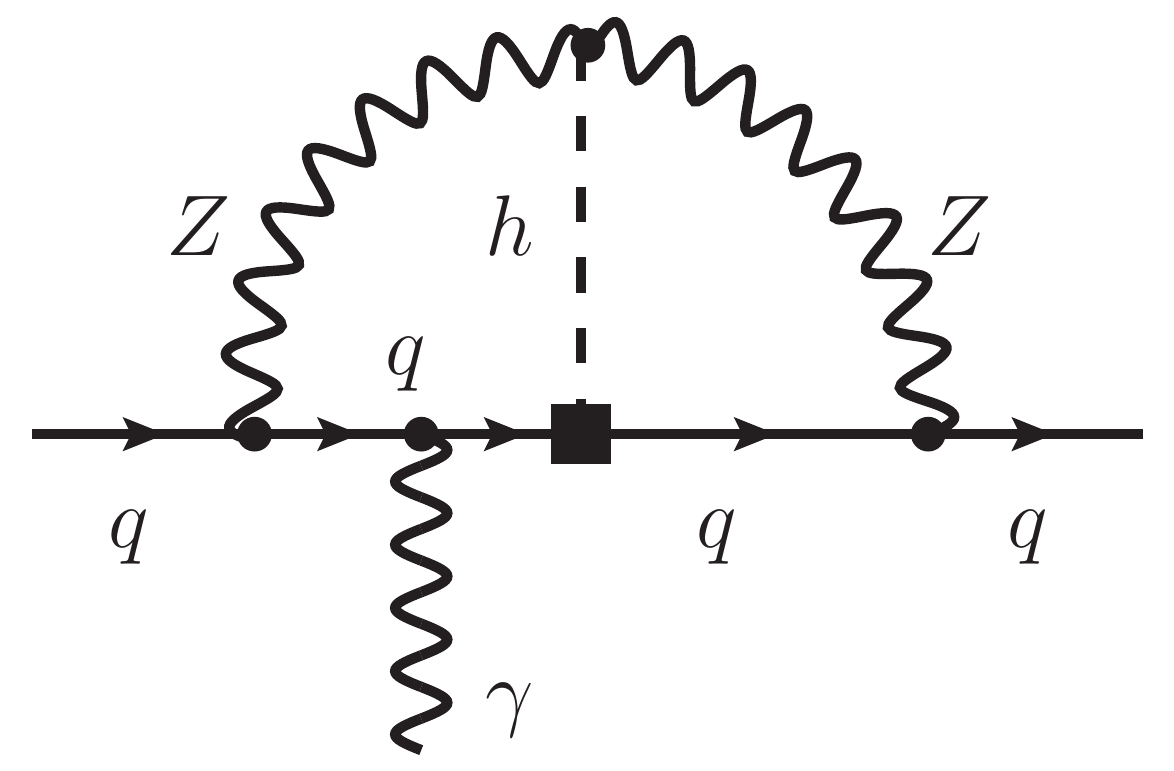}
\end{center}
\caption{Sample two-loop Feynman diagrams with internal gauge bosons,
  inducing an electric dipole moment for the light quark
  $q=u,d,s$. \label{fig:2loopEDM:bosonic}}
\end{figure}

\begin{figure}[t]
\begin{center}
\includegraphics[width=0.3\textwidth]{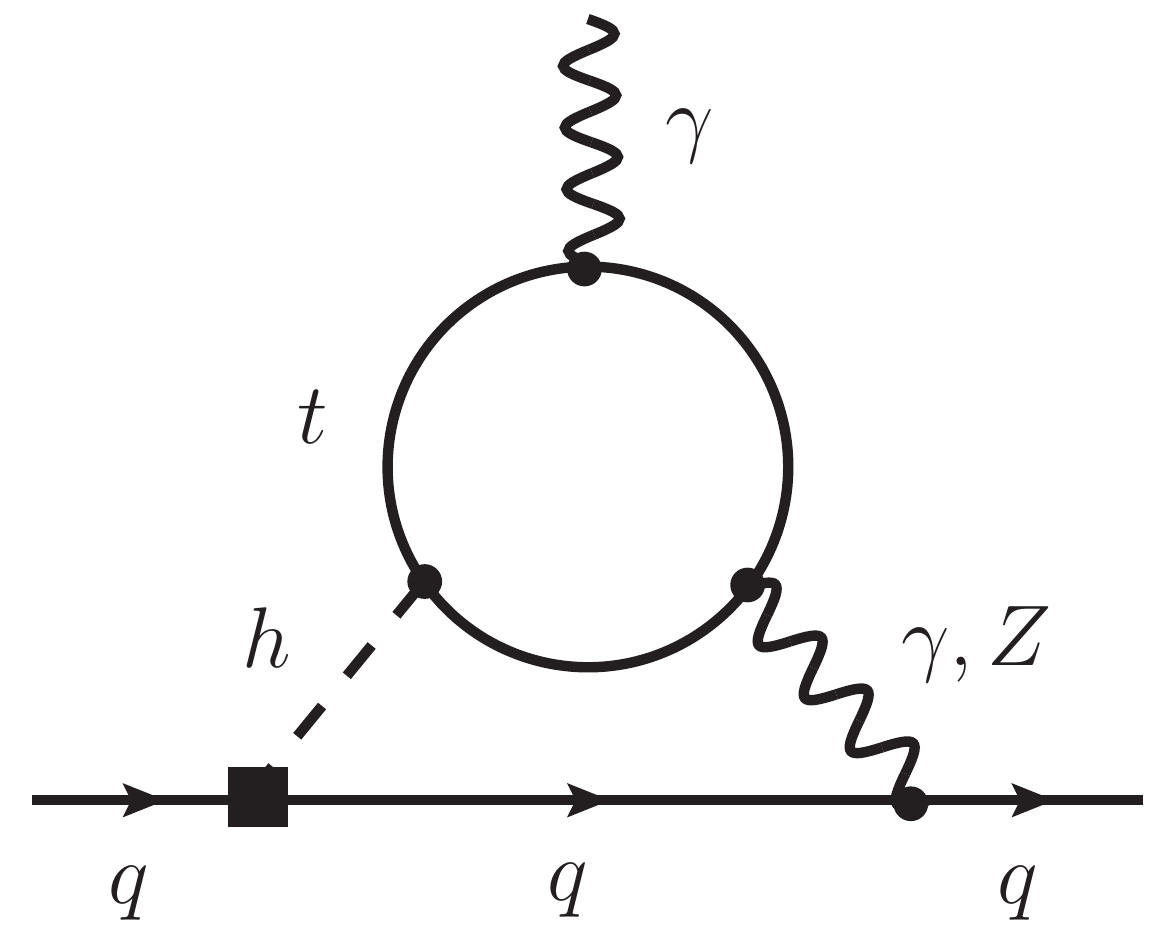}~~~~~
\includegraphics[width=0.3\textwidth]{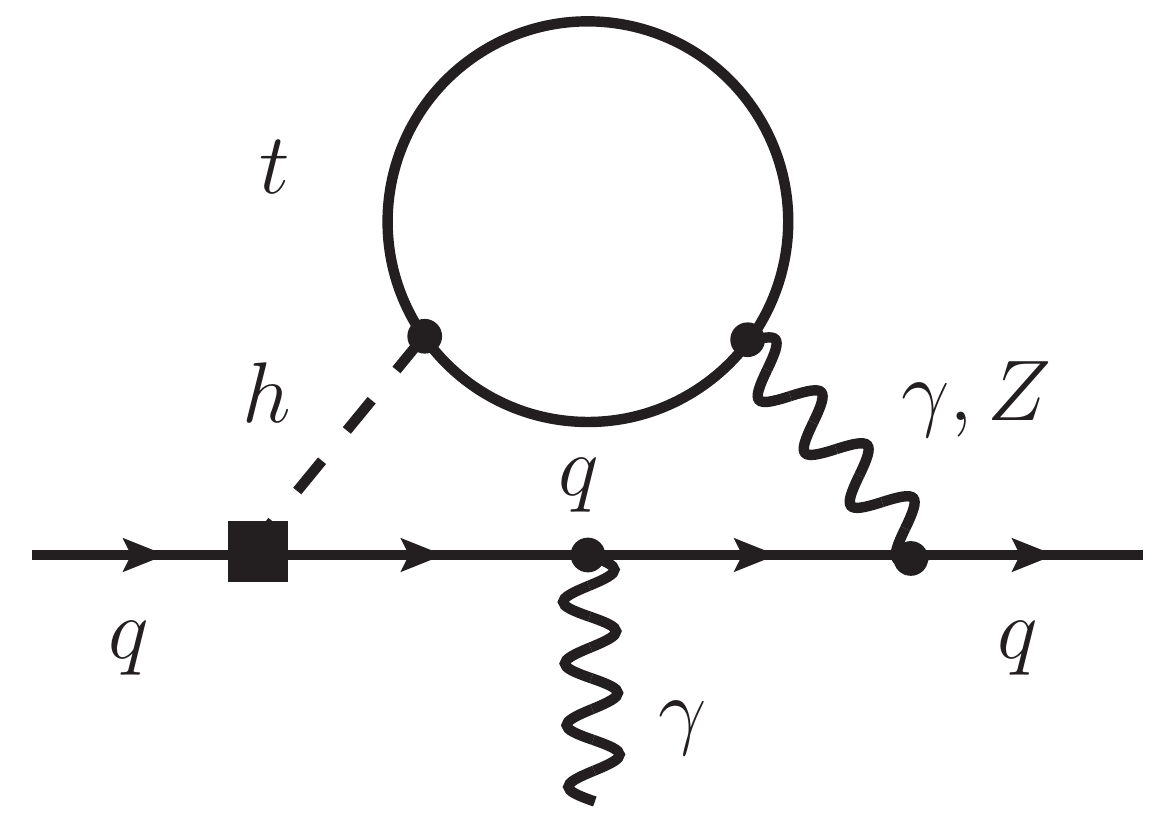}~~~~~
\includegraphics[width=0.3\textwidth]{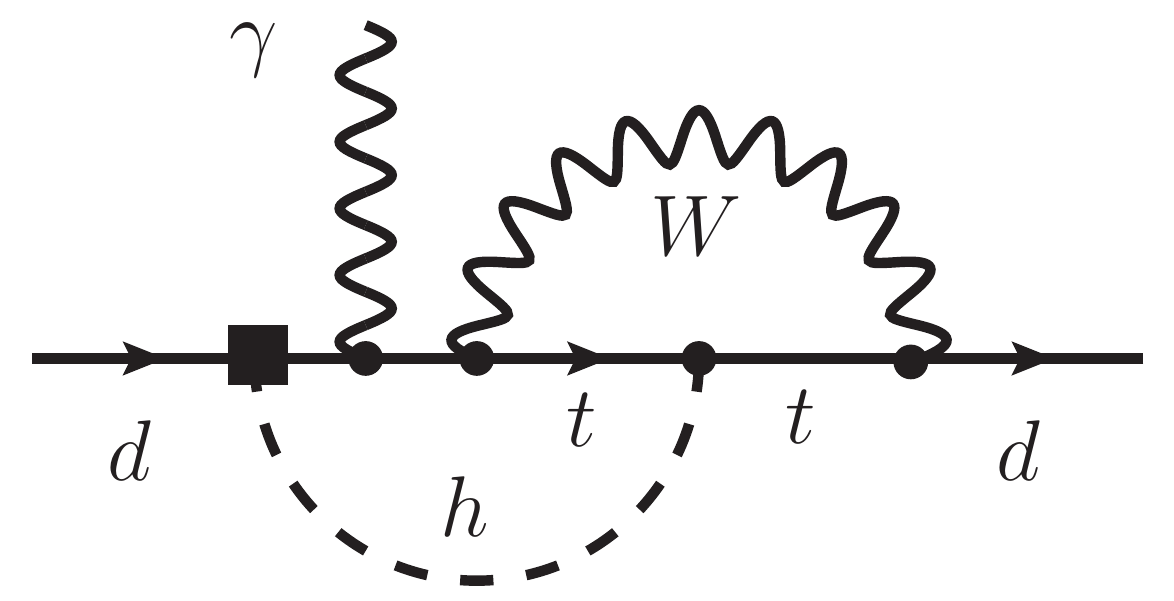}
\end{center}
\caption{Sample two-loop Feynman diagrams with internal top quarks,
  inducing an electric dipole moment for the light quark
  $q=u,d,s$. Left panel: The classic ``Barr--Zee'' diagrams. The
  diagrams with an internal $Z$ boson give a substantial contribution
  for external down quarks. The sum of diagrams of the form shown in
  the middle panel vanishes (the same is true for the corresponding
  bosonic diagrams). Right panel: This class of diagrams with an
  internal top-quark line is parametrically suppressed by small CKM
  matrix elements and has been neglected in our
  calculation. \label{fig:2loopEDM:top}}
\end{figure}

We perform the matching at the weak scale by calculating appropriate
off-shell Greens functions with light external quarks, photons, and
gluons. As pointed out in Ref.~\cite{Barr:1990vd}, the leading
contributions arise from two-loop diagrams in the modified SM (see
Fig.~\ref{fig:2loopEDM:gluonic}-\ref{fig:2loopEDM:top}). In order to
project all our results, we need to include one unphysical operator
that vanishes by the equations of motion (e.o.m.) of the light-quark
fields. It can be chosen as
\begin{equation}\label{eq:eom}
N_1^q = m_q \, \bar q \slashed{D} \slashed{D} i \gamma_5 q \,,
\end{equation}
where the covariant derivative acting on quarks is defined as
\begin{equation} \label{eq:codev}
D_\mu \equiv \partial_\mu - i g_s T^a G_\mu^a  + i e Q_q A_\mu \,,
\end{equation}
with the quark electrical charge $Q_q$.

Many contributions to the initial conditions can be obtained, in
principle, by rescaling the results for the electron
EDM~\cite{Altmannshofer:2015qra} (see also
Ref.~\cite{Gribouk:2005ee}). There are, however, new diagrams that
appear only in the case of light-quark EDMs, since the photon does not
couple to the neutrino in the case of the contributions to the
electron EDM (see Fig.~\ref{fig:2loopEDM:bosonic}, middle panel).
Note that the set of additional ``non-Barr--Zee'' diagrams with an
internal top-quark line (see Fig.~\ref{fig:2loopEDM:top}, right panel)
are suppressed by the CKM factor $|V_{td}|^2 \approx 7 \times 10^{-5}$
and are neglected in our calculation. The corresponding diagrams with
external strange quarks are suppressed by $|V_{ts}|^2 \approx 1.5
\times 10^{-3}$ and are also neglected.

In order to display our analytic results, we decompose the Wilson
coefficients as
\begin{equation} \label{EDM}
\begin{split}
 C_1^q &= C_1^{q,t\gamma} + C_1^{q,tZ} + C_1^{q,W\gamma} + C_1^{q,WZ} + C_1^{q,W} + C_1^{q,Z} \,,\\
 C_2^q &= C_2^{q,tg} + C_2^{q,W} + C_2^{q,Z} \,. 
\end{split}
\end{equation}
The terms labeled by $t\gamma$, $tZ$, and $tg$ denote the
contributions from Barr--Zee-type diagrams containing top-quark loops
and an internal photon, $Z$ boson, or gluon, respectively (see
Fig.~\ref{fig:2loopEDM:gluonic} and Fig.~\ref{fig:2loopEDM:top}, left
panels); the terms labeled by $W\gamma$ and $WZ$ denote corresponding
bosonic diagrams (see Fig.~\ref{fig:2loopEDM:bosonic}, left
panel). The terms labeled $W$ and $Z$ denote ``non-Barr--Zee'' type
diagrams with internal $W$ or $Z$ bosons, respectively (see
Fig.~\ref{fig:2loopEDM:gluonic} and Fig.~\ref{fig:2loopEDM:bosonic},
middle and right panels). By explicit calculation, we find
\begin{equation}
\begin{split}
  C_1^{q,t\gamma} = \frac{\alpha}{12\pi^3} \,\kappa_q \,\kappa_t
  \bigg\{ & \sin(\phi_q) \cos(\phi_t)\,\xth \left[
   \left(2\xth - 1\right) \Phi\left(\frac{1}{4\xth}\right) - 2\left( 2
   + \log \xth \right)
   \right] \\
  & - \cos(\phi_q) \sin(\phi_t)\,\xth \Phi\left(\frac{1}{4\xth}\right) \bigg\}\,,
\end{split}
\end{equation}
where $\xth \equiv m_t^2/M_h^2$. The corresponding diagrams with
gluons instead of photons (see Fig.~\ref{fig:2loopEDM:gluonic}, left
panel) give a contribution to the initial condition of the
chromoelectric dipole operator; we find
\begin{equation}
\begin{split}
  C_2^{q,tg} = \frac{\alpha_s}{32\pi^3} \,\kappa_q \,\kappa_t
  \bigg\{ & \sin(\phi_q) \cos(\phi_t)\,\xth \left[
   \left(2\xth - 1\right) \Phi\left(\frac{1}{4\xth}\right) - 2\left( 2
   + \log \xth \right)
   \right] \\
  & - \cos(\phi_q) \sin(\phi_t)\,\xth \Phi\left(\frac{1}{4\xth}\right) \bigg\}\,.
\end{split}
\end{equation}
For the top-loop diagrams with internal $Z$ bosons we
obtain\footnote{Note that, in the case of top-quark CP violation, the
  off-shell Greens function is finite only after inclusion of an
  appropriate counterterm to cancel the divergence in the one-loop
  mixing of the Higgs and the $Z$ boson. The divergent part of the
  amplitude projects entirely on the e.o.m.-vanishing
  operator~\eqref{eq:eom}, so that the physical Wilson coefficient
  $C_1^{q,tZ}$ is still renormalization-scheme independent.}
\begin{equation}
\begin{split}
	C_1^{q,tZ} & = \frac{\alpha}{384 Q_q \pi^3 s_w^2 c_w^2}
 \kappa_q \kappa_t \left( 8 s_w^2 - 3 \right) \left(
 4 Q_q s_w^2 \mp 1 \right) 
  \frac{\xth\xtz}{\xth - \xtz} \\[0.5em]
 & \quad \times \bigg\{ \sin(\phi_q) \cos(\phi_t) \bigg[\left(1 - 2\xth \right)
    \Phi\left(\frac{1}{4\xth}\right) \\[0.5em]
    & \hspace{10em} - 2\log\xhz  - \left(1 - 2\xtz \right)
   \Phi\left(\frac{1}{4\xtz}\right)\bigg]\\[0.5em]
  & \hspace{2em} + \cos(\phi_q) \sin(\phi_t) \bigg[
   \Phi\left(\frac{1}{4\xth}\right) - \Phi\left(\frac{1}{4\xtz}\right)\bigg] \bigg\}\,,
\end{split}
\end{equation}
where $\xtz \equiv m_t^2/M_Z^2$ and $\xhz \equiv M_h^2/M_Z^2$. Here
and in the following, the upper sign corresponds to the up quark
($q=u$) and the lower sign to the down quarks ($q=d,s$). Moreover, in
the remainder of this work we will assume the SM values $\kappa_t = 1$
and $\phi_t = 0$ for the top-quark couplings. For the bosonic
Barr--Zee-type diagrams we find
\begin{equation}
\begin{split}
 C_1^{q,W\gamma} = & \frac{\alpha}{64\pi^3} \sqrt{2} \,
   \kappa_q \sin(\phi_q)\bigg[ \left( 1 + 6\xwh \right) \left( 2 +
   \log \xwh \right) \\ & \hspace{7em} + \left(7 - 6 \xwh \right)
   \xwh \Phi\left(\frac{1}{4\xwh}\right) \bigg] \,, 
\end{split}
\end{equation}
where $\xwh \equiv M_W^2/M_h^2$, and 
\begin{equation}
\begin{split}
 C_1^{q,WZ} = & - \frac{\alpha}{512 \pi^3 s_w^2} \sqrt{2} \,
   \kappa_q \sin(\phi_q) (\xzh - 1)^{-1} \bigg( 4 s_w^2 \mp \frac{1}{Q_q} \bigg) \\[0.5em]
   & \quad \times \bigg[ \big[3 - 2 \xwh + 2 c_w^2 (6 \xwh - 7)\big]
     \xzh \Phi\left(\frac{1}{4\xwh}\right) \\[0.5em]
   & \qquad + \big[1 + 18 \xwh - 2 c_w^2 (1 + 6 \xwh) - 4 \xzh\big]
     \Phi\left(\frac{1}{4c_w^2}\right) \\[0.5em]
   & \qquad +  \bigg(\frac{1}{c_w^2} + 2 \xzh - 2 (1 + 6 \xwh)\bigg) \log \xzh \bigg]\,,
\end{split}
\end{equation}
where $\xzh \equiv M_Z^2/M_h^2$. The contributions to the electric and
chromoelectric dipoles of diagrams with internal $Z$ bosons are equal;
we find
\begin{equation}
\begin{split}
 C_1^{q,Z} = C_2^{q,Z} = & \frac{\alpha}{9216 \pi^3 s_w^2 c_w^2} \sqrt{2} \,
 \kappa_q \sin(\phi_q) \xzh^2 \bigg[ 2 \big( 1 \mp 4 Q_q s_w^2 + 8 Q_q^2 s_w^4 \big) \\[0.5em]
   & \quad \times \bigg( 24 \xhz - 6 \xhz^2 - \big(4 + 3 \xhz \big) \pi^2\\[0.5em]
   & \qquad \quad + \big(24 \xhz + 6 \xhz^2 - 3 \xhz^3 \big) \Phi\left(\frac{\xhz}{4}\right) \\[0.5em]
   & \qquad \quad + \big(24 + 18 \xhz - 6 \xhz^3 \big) \text{Li}_2 (1 - \xhz)\\[0.5em]
   & \qquad \quad - \big(24 \xhz + 6 \xhz^2\big) \log \xhz \bigg) \\[0.5em]
   & - 12 s_w^2 \big(2 Q_q^2 s_w^2 \mp Q_q \big) \xhz^3 \\[0.5em]
   & \quad \times \bigg( 6 - \big(4 \xhz - \xhz^2 \big) \pi^2
   + \big( 24 - 18 \xhz + 3 \xhz^2 \big)  \Phi\left(\frac{\xhz}{4}\right) \\[0.5em]
   & \qquad \quad + \big( 12 - 48 \xhz - 12 \xhz^2 \big) \text{Li}_2 (1 - \xhz) \\[0.5em]
   & \qquad \quad + 12 \log \xhz - \big( 12 \xhz - 3 \xhz^2 \big) \log^2 \xhz \bigg) \bigg]\,.
\end{split}
\end{equation}
The corresponding contributions with internal $W$ loops are
\begin{equation}\label{eq:C1qW}
\begin{split}
 C_1^{q,W} = & \pm \frac{\alpha}{2304 \pi^3 Q_q s_w^2} \sqrt{2} \,
 \kappa_q \sin(\phi_q) \xhw \\[0.5em]
 & \times \bigg[ \pm Q_q \big( 24 \xwh^2 - 6 \xwh \big)
   - 18 \xwh \mp Q_q \big( 3 \xwh^2 + 4 \xwh^3 \big) \pi^2 \\[0.5em]
   & \qquad + \Big( 18 \big(\xwh - \xwh^2 \big)
   - 9 \pm 3 Q_q \big( 8 \xwh^2 + 2 \xwh - 1 \big) \Big) \Phi\left(\frac{1}{4\xwh}\right) \\[0.5em]
   & \qquad + \Big( 18 \pm 3 Q_q \big( 1 - 4 \xwh^3 - 3 \xwh^2 \big) \Big)
   \Big(2\text{Li}_2 (1 - \xwh) + \log^2 \xwh\Big)\\[0.5em]
   & \qquad + \Big( 18 \xwh \pm Q_q \big( 6 \xwh + 24 \xwh^2 \big) \Big) \log \xwh \bigg]\,
\end{split}
\end{equation}
and
\begin{equation}\label{eq:C2qW}
\begin{split}
 C_2^{q,W} = & \frac{\alpha}{2304 \pi^3 s_w^2} \sqrt{2} \,
 \kappa_q \sin(\phi_q) \xhw \\[0.5em]
 & \times \bigg[ \big( 24 \xwh^2 - 6 \xwh \big)
   - \big( 3 \xwh^2 + 4 \xwh^3 \big) \pi^2 \\[0.5em]
   & \qquad + 3 \big( 8 \xwh^2 + 2 \xwh - 1 \big) \Phi\left(\frac{1}{4\xwh}\right) \\[0.5em]
   & \qquad + 3 \big( 1 - 4 \xwh^3 - 3 \xwh^2 \big)
   \Big(2\text{Li}_2 (1 - \xwh) + \log^2 \xwh\Big)\\[0.5em]
   & \qquad + \big( 6 \xwh + 24 \xwh^2 \big) \log \xwh \bigg]\,,
\end{split}
\end{equation}
where $\xhw \equiv M_h^2/M_W^2$. In Eqs.~\eqref{eq:C1qW}
and~\eqref{eq:C2qW} we suppressed the explicit dependence on the CKM
factors. For $q=u$, these two results should be multiplied by
$|V_{ud}|^2 + |V_{us}|^2 + |V_{ub}|^2$, for $q=d$, by $|V_{ud}|^2 +
|V_{cd}|^2$, and for $q=s$, by $|V_{us}|^2 + |V_{cs}|^2$. In all
cases, the squared CKM matrix elements sum to unity to a very good
approximation. Note that we neglected the tiny contributions of
diagrams with internal top quarks.

To simplify the above expressions we defined $c_w=M_W/M_Z$ and $s_w =
\sqrt{1-c_w^2}$. The function $\Phi(z)$ is given
by~\cite{Davydychev:1992mt}
\begin{equation}
\begin{split}
  \Phi(z) & = 4 \bigg( \frac{z}{1-z} \bigg)^{1/2} \text{Cl}_2 \big(2
  \arcsin(z^{1/2})\big) \,, \\[0.5em]
  \text{Cl}_2 (\theta) & = -
  \int_0^\theta dx \log |2 \sin (x/2)| \,,
\end{split}
\end{equation}
for $z<1$ and by
\begin{equation}
\begin{split}
  \Phi(z) & = \bigg( \frac{z}{z-1} \bigg)^{1/2} \bigg\{ -4 \text{Li}_2
  (\xi) + 2 \log^2 \xi - \log^2 (4z) + \frac{\pi^2}{3} \bigg\} \,, \\[0.5em]
  \xi & = \frac{1 - \big(\frac{z-1}{z}\big)^{1/2}}{2} \,,
\end{split}
\end{equation}
for $z>1$, where ${\rm Li}_2 (x) = -\int_0^x du \, \ln (1-u)/u$ is the
usual dilogarithm.

The actual calculation was performed in two independent setups, both
based on self-written \texttt{FORM}~\cite{Vermaseren:2000nd}
routines. The amplitudes were generated using
\texttt{QGRAF}~\cite{Nogueira:1991ex} and
\texttt{FeynArts}~\cite{Hahn:2000kx}, respectively, using the Feynman
rules in background-field gauge from Ref.~\cite{Denner:1994xt}. Both
setups implement the two-loop recursion presented in
Refs.~\cite{Davydychev:1992mt, Bobeth:1999mk}. Needless to say that
the two calculations yield identical results.

Having obtained the initial conditions of the Wilson coefficients for
the operators in Eq.~\eqref{eq:Leff}, we now use the one-loop
renormalization group (RG) equations to evolve the Wilson coefficients
from the weak scale $\mu_\text{ew}$ down to the hadronic scale
$\mu_\text{had} = 2\,$GeV, integrating out the bottom and charm quarks
at their respective thresholds. This procedure automatically sums the
large logarithms $\log (M_h/\mu_\text{had})$ to leading-logarithmic
(LL) order. We follow the standard procedure and conventions described
in Ref.~\cite{Buchalla:1995vs}.

Focusing on the light quarks $q=u,d,s$ only, the evolution from the
weak scale to the hadronic scale is given by the RG equation
\begin{equation}\label{eq:RGE}
\mu\frac{d}{d\mu} C(\mu) = \gamma^{T} C(\mu) \,,
\end{equation}
where $C(\mu) \equiv (C_{1}^q(\mu), C_{2}^q(\mu))^T$, and the
anomalous dimension matrix is given, to leading order,
by~\cite{Shifman:1976de, Hisano:2012cc}
\begin{align}
\gamma = \frac{\alpha_s}{4\pi}
\begin{pmatrix}
\frac{32}{3} & 0\\[0.5em]
\frac{32}{3} & \frac{28}{3}
\end{pmatrix}\,.
\end{align}
Note that operators with different quark flavors do not mix at
one-loop order. The contributions to the low-energy effective
Lagrangian~\eqref{eq:LeffN} are then given in terms of the Wilson
coefficients at $\mu_\text{had} = 2\,$GeV in the three-flavor
effective theory by the relations
\begin{equation} \label{eq:ddw}
\begin{split}
	d_q (\mu) & = \sqrt{2} G_F \, Q_q e \, m_q \, C_{1}^q (\mu) \,, \\[0.5em]
	\tilde d_q (\mu) & = - \sqrt{2} G_F \, m_q \, C_{2}^q (\mu) \,.
\end{split}
\end{equation}

\section{Numerics}\label{sec:num}

In this section, we evaluate our results numerically and study their
impact on the neutron and mercury EDMs. Using experimental bounds, we
derive constraints on the phases of the light-quark Yukawas at the end
of this section. All numerical input parameters in this section are
taken from Ref.~\cite{PDG2018}.

The numerical size of the individual contributions to the initial
conditions of the Wilson coefficients at the weak scale are
\begin{align}\label{eq:C:muew}
	C_1^u (\mu_\text{ew}) &= (-4.5+14.4) \times 10^{-5}\times\kappa_u \sin(\phi_u)  \,;\\
	C_1^q (\mu_\text{ew}) &= (-5.6+20.1) \times 10^{-5}\times \kappa_q \sin(\phi_q)  \,, \quad q=d,s\,,
\end{align}
where we chose $\mu_\text{ew} = M_h = 125.18\,$GeV. The first terms in
the brackets correspond to the top-loop contributions, see
Fig.~\ref{fig:2loopEDM:top}, left panel. The diagrams with internal
photons have been calculated previously, while the diagrams with
internal $Z$ bosons are included here for the first time. The second
terms correspond to the previously unknown bosonic contributions, see
Fig.~\ref{fig:2loopEDM:bosonic}. It is interesting to note that they
enter with the opposite sign and dominate numerically over the
fermionic contributions, similar to the case of the Higgs decay rate
into two photons. For the contribution to the chromoelectric dipole
operator we find
\begin{align}\label{eq:C2:muew}
	C_2^u (\mu_\text{ew}) &= (-22.1+4.2) \times 10^{-5}\times\kappa_u \sin(\phi_u)  \,;\\
	C_2^q (\mu_\text{ew}) &= (-22.1+3.0) \times 10^{-5}\times \kappa_q \sin(\phi_q)  \,, \quad q=d,s\,,
\end{align}

The LL RG evolution of the Wilson coefficients from
$\mu_\text{ew}=M_h$ to $\mu_\text{had} = 2\,$GeV is obtained by
solving Eq.~\eqref{eq:RGE} numerically. We find
\begin{align}\label{eq:C:muhad}
	C_1^u (\mu_\text{had}) &= \kappa_u \sin(\phi_u) (1.16\pm 0.09) \times 10^{-4} \,;\\
	C_1^q (\mu_\text{had}) &= \kappa_q \sin(\phi_q) (1.45 \pm 0.10) \times 10^{-4} \,, \quad q=d,s\,;\\
	C_2^u (\mu_\text{had}) &= -\kappa_u \sin(\phi_u) (1.07 \pm 0.23) \times 10^{-4} \,;\\
	C_2^q (\mu_\text{had}) &= -\kappa_q \sin(\phi_q) (1.14 \pm 0.23) \times 10^{-4} \,, \quad q=d,s\,.
\end{align}
The central values correspond to the values at $\mu_\text{ew} =
M_h$. We estimate the theoretical uncertainty by studying the residual
scale dependence; the quoted errors correspond to half the size of the
interval obtained by varying the electroweak matching scale in the
interval $M_h/2 < \mu_\text{ew} < 2M_h$. By contrast, the dependence
on the scale where the bottom and charm quarks are integrated out is
negligible.

Using Eq.~\eqref{eq:ddw} we find for the resulting quark-level
electric dipole moments
\begin{equation} \label{eq:d:muhad}
\begin{split}
  \frac{d_u}{e}(\mu_\text{had}) &= \kappa_u \sin(\phi_u) (5.5 \pm 0.4) \times 10^{-26}\, \text{cm} \,, \\
  \frac{d_d}{e}(\mu_\text{had}) &= -\kappa_d \sin(\phi_d) (7.4 \pm 0.5) \times 10^{-26}\, \text{cm} \,, \\
  \frac{d_s}{e}(\mu_\text{had}) &= -\kappa_s \sin(\phi_s) (150 \pm 10) \times 10^{-26}\, \text{cm} \,, \\
\end{split}
\end{equation}
and for the chromoelectric dipole moments
\begin{equation} \label{eq:C2:muhad}
\begin{split}
\tilde d_u(\mu_\text{had}) = \kappa_u \sin(\phi_u) (7.7 \pm 1.6) \times 10^{-26} \, \text{cm}\,, \\
\tilde d_d(\mu_\text{had}) = \kappa_d \sin(\phi_d) (17.4 \pm 3.5) \times 10^{-26} \, \text{cm} \,.
\end{split}
\end{equation}

Assuming a Peccei--Quinn-type solution to the strong CP problem we can
now derive constraints on the modified Yukawa couplings from the
experimental bound on the neutron and mercury EDMs. The contributions
to the neutron EDM are
\begin{equation}
  \frac{d_n}{e} = (1.1 \pm 0.55) (\tilde d_d + 0.5 \tilde d_u) + \Big(
  g_T^u \frac{d_u}{e} + g_T^d \frac{d_d}{e} + g_T^s \frac{d_s}{e}
  \Big)
\end{equation}
where we use the matrix elements of the electric dipole operator
parameterized by $g_T^u = -0.204(11)(10)$, $g_T^d = 0.784(28)(10)$,
$g_T^s = -0.0027(16)$. These values are calculated using lattice QCD
and are converted to the MS-bar scheme at 2\,GeV~\cite{Gupta:2018lvp}
(see also Refs.~\cite{Bhattacharya:2015esa, Bhattacharya:2015wna,
  Yamanaka:2018uud, Alexandrou:2017qyt}). The matrix elements of the
chromoelectric dipole operator are estimated using QCD sum rules and
chiral techniques~\cite{Pospelov:2005pr, Engel:2013lsa}. For prospects
on lattice calculations for the latter, see
Refs.~\cite{Bhattacharya:2015rsa, Bhattacharya:2016rrc}. The
experimental 90\% CL exclusion bound $|d_n| < 2.9 \times 10^{-26} \, e
\, \text{cm}$ obtained in Ref.~\cite{Baker:2006ts} implies the 90\% CL
limits
\begin{equation}
 \kappa_u |\sin\phi_u| < 0.94 \,, \qquad
 \kappa_d |\sin\phi_d| < 0.22 \,, \qquad
 \kappa_s |\sin\phi_s| < 7.2 \,,
\end{equation}
where we allowed for the presence of a single CP phase at a time.

Other hadronic EDMs give complementary bounds. For instance, the
contribution to the mercury EDM is given by~\cite{Pospelov:2005pr}
\begin{equation}
  \frac{d_\text{Hg}}{e} = - 1.8 \times 10^{-4} \big(4_{-2}^{+8}\big)
  \big( \tilde d_u - \tilde d_d \big)\,.
	\label{eq:Hgprediction}
\end{equation}
Considering again the presence of a single phase at a time, the
current upper experimental 95\% CL bound~\cite{Graner:2016ses}
$|d_\text{Hg}| < 7.4 \times 10^{-30} \, e \, \text{cm}$ translates
into the 90\% CL limits
\begin{equation}
	\kappa_u |\sin\phi_u| < 0.11 \,, \qquad
	\kappa_d |\sin\phi_d| < 0.05 \,.
\end{equation}
We neglected the theoretical uncertainty in all our bounds.

\section{Discussion and Conclusions} \label{sec:conclusions}

In this work, we considered the Standard Model with light-quark Yukawa
couplings modified to include a CP-violating phase, and studied the
constraints on these phases arising from experimental bounds on
hadronic electric dipole moments.

We presented the analytic result of a two-loop matching calculation at
the electroweak scale of the modified Standard Model onto an effective
five-flavor effective theory, and the subsequent leading-logarithmic
renormalization-group evolution down to the low-energy scale where the
hadronic matrix elements are evaluated. 

Employing the most recent experimental bounds on the neutron and
mercury electric dipole moments, we derived strong constraints on the
CP phases of the up and down quarks, of the order of several
percent. The phase of the strange quark, on the other hand, is only
weakly constrained. This situation is likely to change with upcoming
new experiments~\cite{Chupp:2017rkp}.

An interesting observation, first made in Ref.~\cite{Chien:2015xha},
is that the neutron and mercury yield quite complementary constraints
on the up and down Yukawa, due to the specific values of the
low-energy partonic dipole contributions induced by the modified
Yukawa couplings. This complementarity is enhanced upon inclusion of
the full electroweak matching contributions. This can be contrasted
with the observation made recently in Ref.~\cite{Brod:2018pli} that
the mercury system yields much weaker constraints on CP-violating
phases in the bottom and charm Yukawas than the bound on the neutron
electric dipole moment. The reason is that the loop-induced
contributions to the up- and down-quark chromoelectric dipole
operators are nearly universal for modified bottom and charm Yukawas,
while for the case of modified light-quark Yukawas, the
isospin-breaking electroweak contributions are sizeable and break that
degeneracy.

These observations further motivate a future global
analysis~\cite{BCSS2019} of constraints from various hadronic and
atomic electric dipole moments on all Yukawa couplings, with the hope
of being able to disentangle the bounds on many of the different
contributions of potentially new sources of CP violation. This might
eventually bring us one step closer to understanding the baryon
asymmetry of our universe.

\phantomsection
\addcontentsline{toc}{section}{Acknowledgments}
\section*{Acknowledgments}

We thank Michael Ramsey-Musolf for useful discussions and Emmanuel
Stamou for valuable cross checks and pointing out missing
contributions in an earlier version of this manuscript. JB thanks the
Galileo Galilei Institute for Theoretical Physics for hospitality and
the INFN for partial support during the completion of this work.


\phantomsection \addcontentsline{toc}{section}{References}
\bibliography{paper} \bibliographystyle{JHEP}


\end{document}